# Negative tunneling magnetoresistance by canted magnetization in MgO/NiO tunnel barriers

Hyunsoo Yang,[1,*] See-Hun Yang,[2] Dong-Chen Qi,[3] Andrivo Rusydi,[3] Hiroyo Kawai,[4] Mark Saeys,[5] Titus Leo,[6] David J. Smith,[6] and Stuart S. P. Parkin[2,†]

[1] Department of Electrical and Computer Engineering and NUSNNI-NanoCore, National University of Singapore, 117576, Singapore

[2] IBM Almaden Research Center, 650 Harry Road, San Jose, California 95120

[3] NUSNNI-NanoCore and Departments of Physics, Singapore Synchrotron Light Source, National University of Singapore, 117603, Singapore

[4] Institute of Materials Research and Engineering, 3 Research Link, Singapore 117602

[5] Department of Chemical and Biomolecular Engineering, National University of Singapore, 117576, Singapore

[6] School of Materials and Department of Physics, Arizona State University, Tempe, Arizona 85287

The influence of insertion of an ultra-thin NiO layer between the MgO barrier and ferromagnetic electrode in magnetic tunnel junctions has been investigated by measuring the tunneling magnetoresistance and the X-ray magnetic circular dichroism (XMCD). The magnetoresistance shows a high asymmetry with respect to bias voltage, giving rise to a negative value of -16% at 2.8 K. We attribute this to the formation of non-collinear spin structures in the NiO layer as observed by XMCD. The magnetic moments of the interface Ni atoms tilt from the easy axis due to exchange interaction and the tilting angle decreases with increasing the NiO thickness. The experimental observations are further support by non-collinear spin density functional theory.

In nanoscale multilayers in which the ratio between the interface and the bulk is comparable, artificial engineering on the level of individual monolayer (ML) is pivotal, especially, to spintronics. One of the key examples is magnetic tunnel junctions (MTJs), consisting of two ferromagnets (FM) separated by a thin insulating barrier[1, 2]. Recently, very high tunneling magnetoresistance (TMR) values of more than 600% have been reported in MgO MTJs[3-5]. In most MTJs, the magnitude of the TMR decreases with both increasing bias voltage and temperature. It is generally believed that increased contributions from magnons, magnetic impurities, localized trap states, and the modified electronic structure at elevated biases and temperatures are responsible[6].

The role of the FM/MgO interface in the bias voltage dependence of TMR values and signs has been studied [7-11]. It was reported that the development of quantum well states, the modification of the interfacial polarization, the localized resonance state in the minority band of Fe, and the insertion of impurities develop a strongly asymmetric bias voltage dependence of TMR in MgO MTJs [9-13]. However, the detailed correlation between TMR and magnetic/electronic structures of the interface layer has been yet to be understood. The TMR values increase with anneal treatments, likely due to the improved tunnel barrier and interfaces of barrier/ferromagnetic layers. In MTJs the formation of the interfacial oxides cannot be ruled out. As a result, the ultrathin transition metal oxide (e.g. NiO, CoO, or FeO) is present at interfaces, which modifies the magneto-transport and interfacial magnetic/electronic structures. Transition metal oxides such as FeO, $Fe_3O_4$, and $CrO_2$ combined with MgO barrier were studied and the origin of a negative TMR has been ascribed to the effect of interfacial polarization [12-14]. However, no solid experimental proof has been reported and the theoretical models have been limited mostly to collinear magnetic order of the thin transition metal oxides thus far [7, 15].

We report here that the insertion of thin NiO layer in the MgO barrier results in dramatic changes in the TMR, $(R_{AP}-R_P)/R_P$, where $R_P$ and $R_{AP}$ is the differential resistance in the parallel (P) and antiparallel (AP) alignment of two FM, respectively. The magnitude of the TMR is significantly decreased independent of whether these layers are inserted at the MgO interfaces or within the MgO layer. More importantly, the sign of the TMR changes, resulting in a negative TMR value at certain bias voltages. X-ray magnetic circular dichroism (XMCD) measurements reveal that Ni spin moments at the interfacial NiO layer are canted, supported by non-collinear spin density functional theory. The calculated interfacial polarization of the projected spin density of states (DOS) accounts for the negative TMR values.

NiO has a similar simple cubic structure to MgO and grow epitaxially on MgO [16]. Details of the sample preparation are described elsewhere [17]. The samples were not subject to any subsequent annealing treatments in order to prevent any diffusion of the NiO layer. A typical example of the magneto-transport data is shown in Fig. 1(a) measured from 100 MgO/50 Ta/250 $Ir_{22}Mn_{78}$/3 $Co_{49}Fe_{21}B_{30}$/37 $Co_{70}Fe_{30}$/8 Mg/23 MgO/4 NiO/25 $Co_{70}Fe_{30}$/150 $Co_{49}Fe_{21}B_{30}$/100 Ta (thicknesses in Å). A highly asymmetric bias voltage dependence is observed with negative TMR at negative bias voltages < -0.15 V. Negative bias voltages correspond to electrons tunneling from the CoFe/MgO into the NiO/CoFe interface. By changing the barrier order (NiO/MgO), the asymmetric TMR is flipped with respect to the bias voltage in Fig. 1(b), showing that the NiO layer is the origin of TMR inversion. It is remarkable that even a slight shift of the NiO layer into the middle of the barrier (21 Å MgO/4 Å NiO/2.5 Å MgO) ends up with restoration of the symmetry of TMR and the negative TMR disappears in Fig. 1(c), which suggests that electronic/magnetic structure at NiO/CoFe interface plays a key role in asymmetric and negative TMR. When the NiO layer is in the middle of the barrier (12 Å MgO/4 Å NiO/12 Å MgO), the zero bias TMR values changes from 1% at room temperature up to more than 60% at 2.8 K in Fig. 1(d), whereas the MTJ with NiO/CoFe interface shows only twice increase of TMR values at zero bias with a similar temperature change in Fig. 2. This dramatic enhancement of TMR at low bias voltage and low temperature is due to the higher order cotunneling effect across the NiO layer [17, 18].

Figure 2 exhibits the bias voltage dependence of TMR with different barrier thicknesses. Without any NiO layer, the TMR is symmetric and shows a large TMR up to 120% at 2.8 K. Insertion of one ML of NiO (~2 Å) leads to a slightly asymmetric TMR. With increasing the NiO layer (4 Å) a pronounced asymmetric bias dependence of the TMR is observed with a maximum negative TMR of -16% at -0.47 V and 2.8 K. As the thickness of NiO increases above 4 Å, the magnitude of the negative TMR as well as the asymmetry of TMR diminishes. Our observation cannot be simply accounted for with the barrier profile effect [19], since a thicker NiO layer (more asymmetric barrier) would produce higher asymmetry in TMR, which is opposite to what we observe. This implies that the interfacial magnetic and electronic structure in the NiO/CoFe interface is important and should be taken into account.

The interfacial magnetic structures were probed by XMCD in Fig. 3(c). Intriguingly, the Ni $L_{3,2}$ edges exhibits strong XMCD signals with the same sign as that from Co in the FM electrode [16], indicating net magnetic moments of Ni with a component parallel to that of Co atoms. This is surprising because moments on the ideal bulk NiO(001) plane are fully compensated. Moreover, the Ni $L_{3,2}$ edge XMCD signal decreases with increasing NiO layer thickness (2-8 Å). In comparison, no XMCD signal was observed on a 3000 Å thick NiO film without the CoFe, thus showing the indispensible role of FM in inducing the ferromagnetic ordering of interfacial Ni spins, which coincides with the condition for negative TMR (see Fig. 3(c) and Fig. S5 [16]).

In order to understand the exchange coupling mechanism that gives rise to the ferromagnetic XMCD signal of Ni moments in NiO layer, the individual spin and orbit magnetic moment ($m_{spin}$ and $m_{orb}$) in units of $\mu_B$/atom for different NiO thickness were determined from the XMCD and X-ray absorption spectra according to their corresponding sum rules (see Fig. 3(c) and Fig. S4) [16, 20, 21]. The largest spin moment ($m_{spin}^{NiO}$ = 0.496) is deduced for 2 Å and 4 Å NiO thicknesses, and is about 25% of that for NiO bulk crystals ($m_{spin}^{Ni} \approx 1.9$) measured by magnetic X-ray scattering [22]. The significantly reduced spin moment suggests that the Ni moments are aligned almost perpendicular not only to the light propagation direction (***k***) but to the CoFe moments with small component parallel to CoFe magnetic moments. Such perpendicular alignment between the FM and antiferromagnetic (AFM) spins, spin-flop coupling, was theoretically predicted and recently demonstrated at the Co/NiO interface [23-25]. The spin-flop coupling, whose strength depends on the competition between the collinear interfacial exchange interaction of the magnetic moments and the AFM superexchange interaction, leads to slightly canted spin direction of the interfacial NiO layer away from the AFM easy axis. The proposed spin configuration near the interface plane of CoFe/NiO is illustrated in Fig. 3(b), in which the Ni moments are aligned in-plane along the Co <1 $\bar{1}$ 0> direction with a small canting angle $\theta$ [25]. From this model, an average $\theta$ of 15º is estimated for 2 Å and 4 Å NiO, reducing to about 6º for 8 Å NiO shown in Fig. 3(d). The reduced canting angle is due to enhanced effective AFM exchange coupling which in turn suppresses the spin-flop coupling for the thick NiO layer [26].

The effect of the thin NiO layer insertion has been investigated by use of non-collinear spin density functional theory with the Perdew-Burke-Ernzerhof functional (DFT-PBE) as implemented in the Vienna *ab initio* Simulation Package (VASP) [27, 28]. The GGA+U method with an optimized Hubbard U parameter of 4 eV has been used to deal with the localized 3$d$ orbital in NiO [29, 30]. A layer of Co/NiO/MgO with varying NiO thickness was used for the model. Figure 3(a) shows the atomic model with 2 ML NiO. AFM NiO layers were placed on a four-layer bcc Co(100) slab at a 45° angle, and the easy axis of Co electrode is along <110> direction. The top view of the structure and the calculated Ni and interface Co magnetic moments for a Co/2 ML NiO/MgO layer are shown in Fig. 3(b). While the Co magnetic moments at the interface remain close to the <110> direction, the Ni moments at the interface tilt towards the Co moments due to exchange interaction, with a tilting angle ($\theta_1$) of about 14°, in excellent agreement with the experiment. The tilting angle for the second NiO layer ($\theta_2$) changes to -5° due to antiferromagnetic coupling with neighboring Ni atoms. In general, the average tilting angle $\theta$ is calculated to decrease with increasing thickness of NiO as shown in Fig. 3(d), in which the average tilting angles are also compared with those by XMCD. The calculated angles qualitatively agree with the experiment, but some quantitative discrepancy is observed; e.g., the tilting angles obtained from XMCD do not saturate for the NiO layer thicker than 6 Å. This may be possibly due to a non-uniform NiO thickness because the tilting angles would be bigger in the region where NiO is thinner than the nominal value whereas the angles for thicker NiO region would be same to those for the nominal NiO thickness. As a result, non-uniformity of NiO film leads to the enhanced tilted angles. This also accounts for the disagreement between the experiment and the calculation for 2 and 4 Å NiO. In addition, the calculated spin magnetic moment for bulk NiO of 1.6 $\mu_B$ is somewhat smaller than the experimental value of 1.9 $\mu_B$ [22].

To further investigate the impact of the canted interface spin on the electronic properties (spin polarization) and eventually explain the asymmetric and negative TMR phenomenon, the $s$ and $p$ orbital DOS of the NiO layer projected on the <110> direction, the easy axis of the Co electrode, were calculated for the MgO/NiO interface in Fig. 4 (a-d). The $d$ electrons are expected to contribute less to the overall transmission coefficient of the tunnel junction since the $d$ orbitals are more localized and compact, therefore we have focused on $s$ and $p$ orbitals in our analysis. Note that the tunneling process in our MgO barriers is not fully coherent because the devices were not annealed. In non-collinear calculations, the spin quantization axis may be different for each layer, and therefore the tunneling probability across the multilayer depends on both the angle between the spin quantization axes and the spin-dependent tunneling DOS or spin polarization [31]. Non-collinear DFT as implemented in VASP incorporates both effects by calculating the difference in the up and down spin DOS ($N_\uparrow - N_\downarrow$) projected onto the x, y, and z axis for each atom.

The difference between the up and down spin DOS in the <110> direction at the MgO/NiO interface is compared with the spin polarization for the Co electrode in Fig. 4(a-d). The interaction of a thin layer of NiO with the Co electrode produces NiO states near the Fermi level ($E_F$). Because of the spin tilting, those states show a magnetization in the Co <110> direction in Fig. 4(a-d). For 2 and 3 ML NiO, Ni and O $p_z$ states can be found above $E_F$ for the Co/NiO/MgO stack. For thicker NiO films, the NiO band gap gradually opens up and no states are found near $E_F$ for more than 4 layers of NiO [16]. The NiO states above $E_F$ play a critical role in the tunneling transport for negative bias when electrons tunnel from the occupied states of the left Co/MgO interface to the empty states of the right NiO/Co interface shown in Fig. 4(e). Therefore, a thin NiO layer does not behave as an insulator, and the tunneling properties are determined by both the Co DOS and the NiO DOS at NiO/Co interface. The *s* and *p* DOS for the Co electrodes are *positively* spin polarized in the energy range of $E = E_F \pm 1$ eV, except for E ~ -1 eV where the number of minority *p* states exceeds the number of majority *p* states in Fig. 4(a-d). On the other hand, the difference *p* DOS for Ni and O is significantly *negative* in the energy range $E = E_F \pm 1$ eV for 1 ML NiO, and both the *s* and *p* electrons of the interface NiO layer are *negatively* spin polarized above $E_F$ for 2 ML and 3 ML NiO. Consequently, for a negative bias where the electrons tunnel from the the left Co electrode through the MgO barrier into the right NiO/Co electrode in Fig. 4(e), the tunneling probability is expected to be higher for an AP configuration with a few ML of NiO. From the spin polarizations of NiO and Co, a TMR of -36% at -0.73 V can be calculated using a Julliere-type model for 1 ML NiO, in line with the experimental data in Fig. 2 [16]. For a thick NiO layer in Fig. 4(f), the NiO layer acts as another barrier, and the tunneling probability is low for the AP configuration since the left and right Co electrodes are oppositely polarized. The TMR is, therefore, expected to become negative as observed experimentally for thinner NiO layers in Fig. 2, which gradually disappears as the thickness of NiO increases due to the formation of a band gap, thus there are less negatively polarized states in NiO.

In summary, we find a highly asymmetric bias voltage dependence of TMR in CoFe/MgO/NiO/CoFe MTJs. The observed negative TMR at certain bias voltage conditions cannot be accounted for without considering the interfacial magnetic and electronic structures of tunnel barriers. XMCD measurements and non-collinear spin density functional theory identified a canted non-collinear spin configuration at the interface due to the insertion of NiO at the MgO/FM interface, which is responsible for the highly asymmetric and negative TMR for a certain bias voltage region.

This work was partially supported by the Singapore MOE2008-T2-1-105, NRF-CRP 4-2008-06, NRF-CRP002-024, YIA, cross faculty grant, and FRC.

[*] Electronic address: eleyang@nus.edu.sg
[†] Electronic address: parkin@almaden.ibm.com

Fig. 1. The voltage dependence of TMR from MTJs of CoFe/23 Å MgO/4 Å NiO/CoFe (a), CoFe/4 Å NiO/25 Å MgO/CoFe (b), CoFe/21 Å MgO/4 Å NiO/2.5 Å MgO/CoFe (c), and CoFe/12 Å MgO/4 Å NiO/12 Å MgO/CoFe (d). For clarity, the data at 290 K in (d) are multiplied by 10. The junctions are of 240×80 μm$^2$ in size and $R_P$ ranges ~1-5 kΩ at zero bias and 2.8 K.

Fig. 2. Bias dependence of TMR from an MTJ with a structure of CoFe/x MgO/y NiO/CoFe (x+y = 27 Å) at 2.8 K (a) and 290 K (b).

Fig. 3. (a) A Co/2 ML NiO/MgO stack used in simulations. (b) Top view of the Co interface (IF) and Ni atoms for the structure in (a). The MgO layer is not shown. The unit cell is indicated by the rectangle, and θ is the tilting angle between the Ni magnetic moment and the <1 $\bar{1}$ 0> direction. (c) Dichroism for Ni from samples of different NiO thicknesses from CoFe/x MgO/y NiO/CoFe (x+y = 27 Å) and CoFe/27 Å MgO/3000 Å NiO. The integration of one representative XMCD spectrum (y = 4 Å ) is shown in dotted line, in which $p_{NiO}$ and $q_{NiO}$ are the integration values over the $L_3$ range and the whole $L_3+L_2$ range, respectively. (d) Calculated change in the tilting angle θ with NiO thickness for each layer (bars) and in the average angle θ with NiO thickness (open symbol) with the experimental XMCD data (filled symbol). The first NiO layer is next to Co and the last one is next to MgO when there are more than 1 NiO ML.

Fig. 4. (a-d) Difference between up and down spin DOS of the *s* and *p* orbitals in the <110> direction for Ni and O at the NiO/MgO interface with 1 to 4 ML of NiO, and for the Co electrode. The Fermi level is at zero energy and is indicated by the dotted line. (e-f) Potential diagrams for a stack with thin (e) and thick (f) NiO layer.

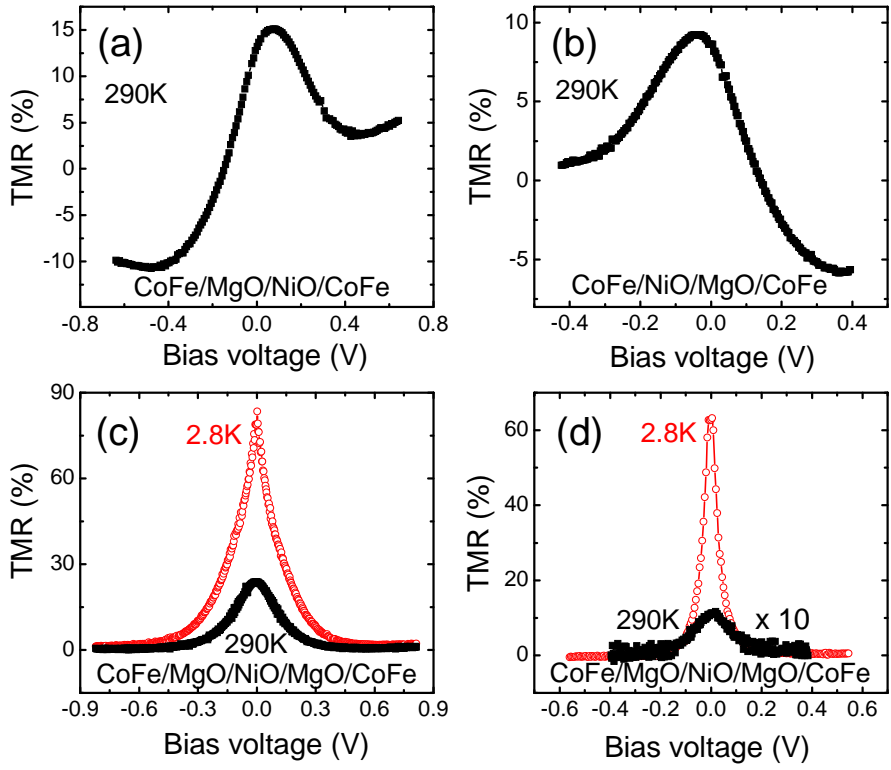

Figure 1.

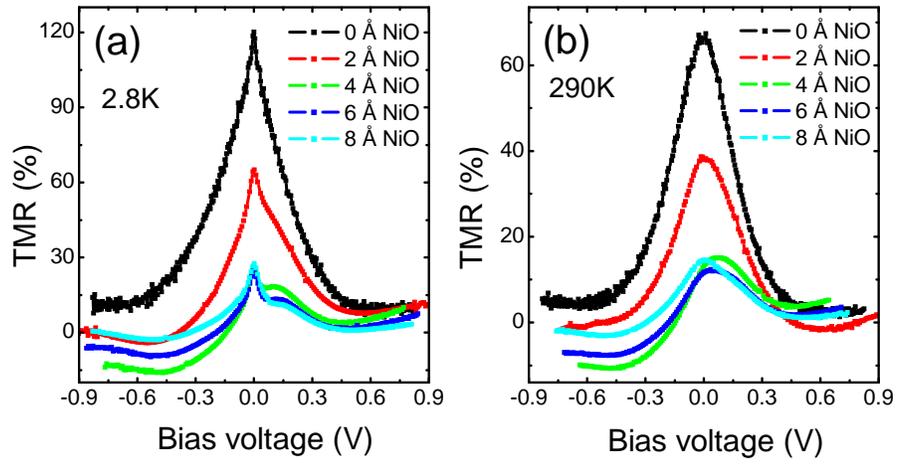

Figure 2.

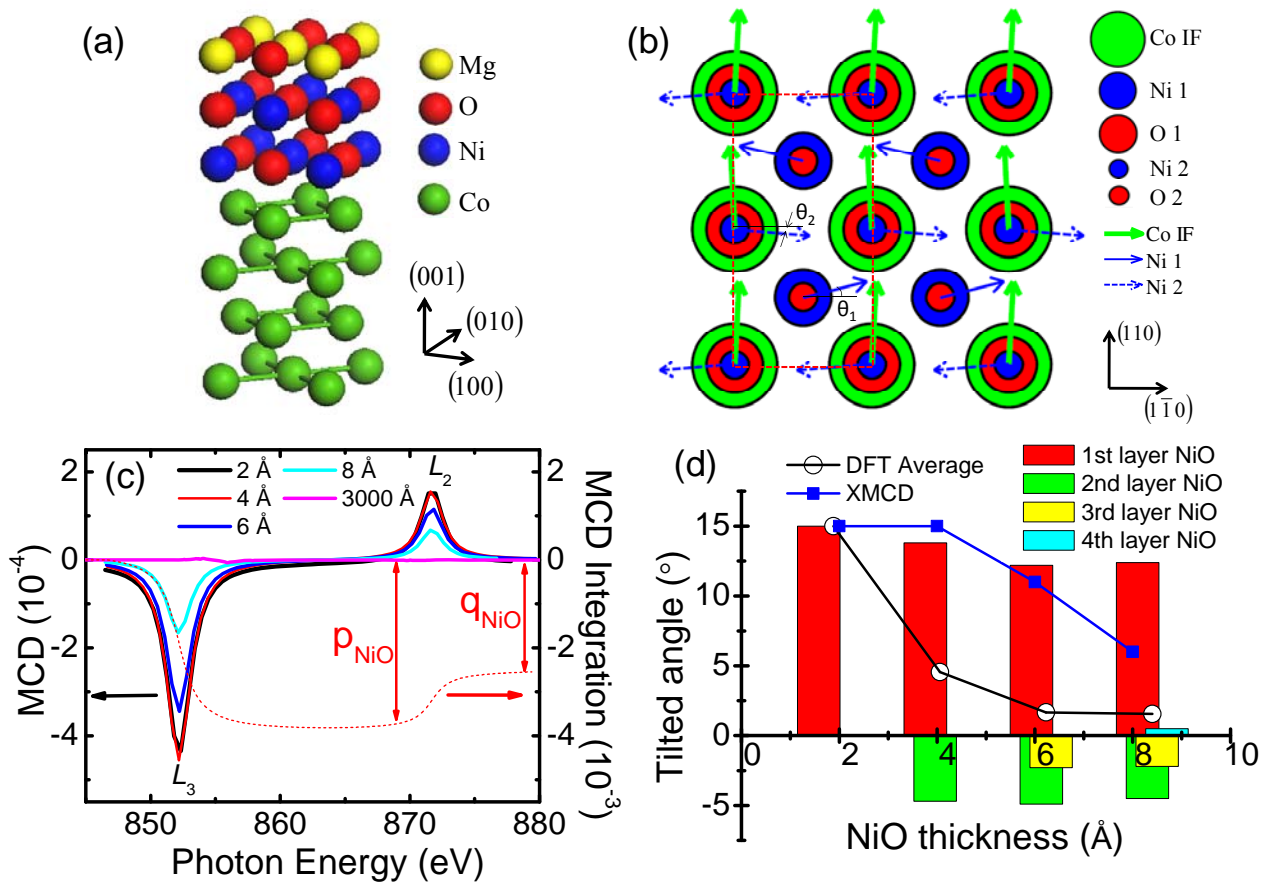

Figure 3.

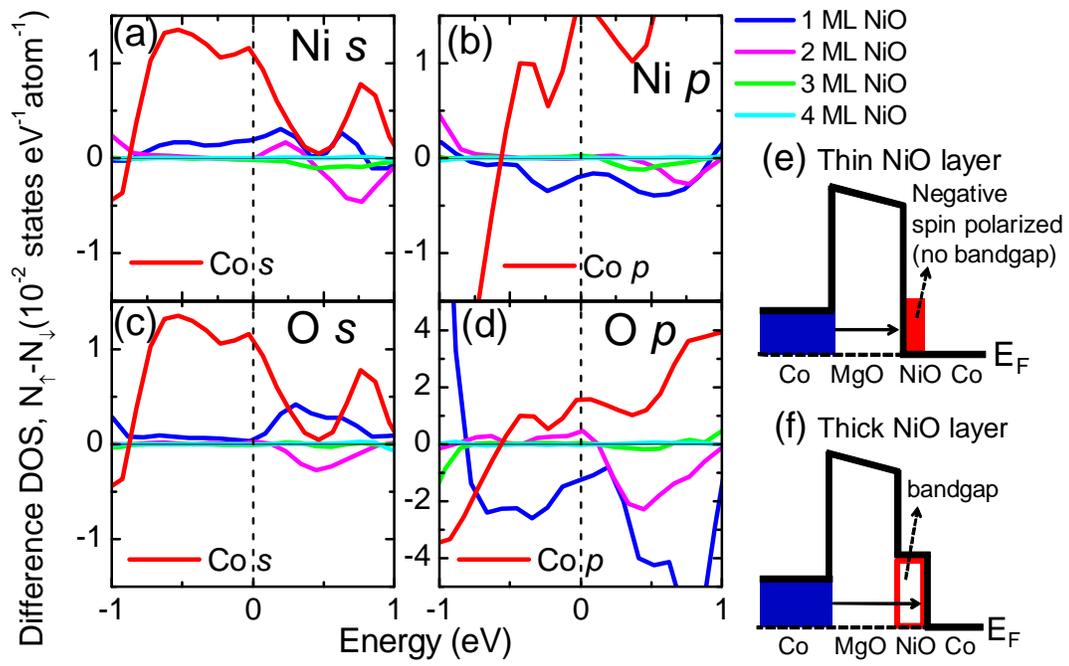

Figure 4.